\documentclass{aastex63}
\usepackage{amsmath}
\usepackage{mathtools}
\usepackage[toc,page]{appendix}
\usepackage[caption = false]{subfig}
\usepackage{natbib}
\begin{document}
\title{Validating inversions for toroidal flows using normal-mode coupling}

\author{Prasad Subramanian \& Shravan Hanasoge}
\affiliation{Department of Astronomy $\&$ Astrophysics, Tata Institute of Fundamental Research, Mumbai 400005, India; prasad.subramanian@tifr.res.in}

\begin{abstract}
Normal-mode coupling is a helioseismic technique that uses measurements of mode eigenfunctions to infer interior structure of the Sun. This technique has led to insights into the evolution and structure of toroidal flows in the solar interior. Here, we validate an inversion algorithm for normal-mode coupling by generating synthetic seismic measurements associated with input flows and comparing the input and inverted velocities. We study four different cases of input toroidal flows and compute synthetics that take into account the partial visibility of the Sun. We invert the synthetics using Subtractive Optimally Localized Averages (SOLA) and also try to mitigate the systematics of mode leakage. We demonstrate that, ultimately, inversions are only as good as the model we assume for the correlation between flow velocities.
\end{abstract}

\section{Introduction} \label{sec:intro}
Convection, the mode of heat transport in the outer one-third of the solar radius, is widely accepted as the driver of large-scale dynamics of the Sun, acting as a transporter of fluid angular momentum and redistributing it, giving rise to differential rotation, meridional circulation, solar subsurface weather and a host of associated flows \citep{Miesch}. Convection is also responsible for exciting acoustic oscillations that resonate between the surface and the solar interior, carrying information about the medium in which they propagate \citep[e.g.][]{CD2000}. Solar magnetic fields arise out of and are sustained by convective flows and other magneto-hydrodynamic processes; on a larger-scale, convection interacts nonlinearly with magnetism possibly giving rise to the 11-year magnetic cycle \citep{Cat1999, BandB2017}. Convection is thought to significantly influence the dynamics of the solar chromosphere \citep{SandL1964} and corona, which in turn plays a role in controlling space weather by indirectly modulating eruptive events, such as solar flares and coronal mass ejections \citep[sudden outbursts of energy from the photosphere;][]{SandN2009,DandT2004,Sch1997}. 

Helioseismology serves as a powerful tool with which to investigate convective flow \citep[][]{Basu1999,Duvall2000,Howe2006,Woodard2007} since the properties and features of a medium are encoded in the modes \citep[e.g.,][]{CD2000}. Studies conducted using helioseismology have suggested that convective velocities are substantially smaller than those predicted by theory and simulations \citep{Han2012,Han2016}, although there is some controversy regarding this \citep{Greer2015}. \citet{HanSciA2020} analysed of 8 years of global mode time-series from the Helioseismic Magnetic Imager \citep[HMI;][]{HMI2012} instrument, on-board the Solar Dynamics Observatory (SDO) and found that power-spectra obtained from simulations of convective flow \citep{Hotta2016} and those from seismic analyses of observations did not compare well. There has also been quite a bit of work on numerical simulations of convection \citep[][]{Zhao2007,Hartlep2013,Miesch2015}. 

Helioseismology has helped significantly improve our understanding of the Sun's structure \citep[][]{CD1991,Gough1996,Basu1997,Lindsey2000,DiMauro2002}. Helioseismic investigations have yielded insights into differential rotation \citep[][]{Schou1998} in the radial \citep[][]{Deubner1979,Duvall1984} and in the latitudinal \citep[][]{Brown1985,Birch1998} directions; \citep[for a broad overview, see][]{Thompson2003,Howe2009}. Our appreciation for meridional circulation, although still an active area of research, has been greatly improved through seismic investigation \citep[][]{Hathaway1996,Giles1997,Hathaway2012,Zhao2013,Jackiewicz2015,Rajaguru2015, Basu2010,Gough2010,Schad2013}. The mapping of thermal variations in the Sun \citep[][]{CD1989,Basu1997b,Baturin2000}, specifically sunspots  \citep[][]{Zhao2003b,Gizon2009}. \citet{Kosovichev2000}, \citet{Gizon2010}, and \citet{Basu2016} has benefited from local and global helioseismology.
 
Helioseismic inquiry typically proceeds on two fronts, commonly undertaken in lockstep. In a forward calculation, one constructs realistic observables by making use of a wave theory that connects internal model properties to the observable. The technique of normal-mode coupling, detailed in, e.g., \citet{LandR1992}, allows for devising a forward model to construct observables that encapsulate the physics that influence oscillations. In the inverse problem, one infers the internal properties of the medium encoded in the observables.

Normal-mode coupling is a helioseismic tool that has found currency in recent times. As a tool that is frequently used to devise forward models, it expresses the solar oscillation wavefield using a complete, orthonormal basis of oscillation eigenfunctions derived for a reference model. The reference basis is obtained by solving an eigenvalue-eigenfunction equation that describes oscillations in the reference model. Thus solar eigenfunctions are in a `coupled state' with respect to the reference eigenfunctions \citep{Han2017}. A common reference model used for such a purpose in helioseismology is Model S \citet{CD1996}, in which the Sun is treated as spherically symmetric, non-rotating, non-magnetic, isotropic and temporally stationary and the acoustic oscillations are treated as adiabatic \citep{LandR1992}. The model is assumed to be sufficiently close to the Sun, allowing us to invoke perturbation theory to represent the Sun as a small deviation from the reference model. 

The use of normal-mode coupling in helioseismology goes back to \citet{Wood1989}, who described a method to express oscillation eigenfunctions of the rotating Sun as superpositions over eigenfunctions of a  non-rotating model. Subsequently, \citet{LandR1992} calculated how global convection can affect the oscillation eigenfunctions and eigenfrequencies and presented a formalism with which mode-coupling theory could be applied computationally. \citet{Wood2016} used mode-coupling theory to interpret correlations of Michelson Doppler Imager (MDI) data \citep{MDI1995} of spherical harmonic time series and concluded that toroidal flow velocities were of amplitude $\approx30$~m/s. \citet{Han2017} reworked the algebra of \citet{LandR1992}, elucidating the connection to helioseismic measurements more clearly. Accounting for the limited visibility ($<2\pi$Sr) of the Sun, \citet{Han2018} derived more realistic mode-coupling sensitivities to toroidal flows in the interior. 

This paper details the results of synthetic tests where simulations of convective flow are used as input to a forward model derived from normal-mode coupling. This work investigates the robustness of mode-coupling as an investigative technique in inferring convective flows over a range of scales.

Research in mode coupling has seen a flurry of activity in recent years \citep[e.g.,][]{Roth2011,Roth2020,Sri2020,HandM2019,MandH2020,Han2017_2}. The first paper on the topic of mode-coupling \citep[][]{Wood1989} focused on inferring the latitudinal variation of differential rotation, a work that was extended \citep[][]{Vor2007,Vor2011} to include radial variations and influence of higher order effects. \citet{Roth2011} and \citet{Roth2020} discuss perturbation of $p$-mode eigenfunctions due to meridional flow and differential rotation, respectively. \citet{HandM2019} state that Rossby-modes \citep{Loptein2018} can be used as a comparative means between different helioseismic methods such as time-distance \citep{Duvall1993}, ring diagrams \citep{Hill1988} and mode-coupling. \citet{MandH2020} extend their work to model systematic effects of leakage on mode-coupling measurements obtained from HMI and MDI.

Before briefly outlining the inverse problem, we highlight a few findings from \citet{HanSciA2020} - which we refer to as H20 in this article - that serve to situate the work in this paper in proper context. H20 explain the importance of contrasting global convective simulations with observations of convective flow because it helps in developing a better understanding of large-scale phenomena driven by turbulence in the convection zone. They find that toroidal flows, a part of the total flow that only contains horizontal components, grows in power as spatial wave number and temporal frequency increase. This is in direct contrast to simulations of solar convection, which show that power in toroidal flows decreases with spatial wave number and temporal frequency. They also note that observed flows are confined to the equatorial region and weak at high latitudes, opposite to the trend seen in simulations. Although H20 performed inversions, they did not thoroughly investigate the reliability of their algorithms. It is important to exclude inversion errors when interpreting the inferences and determine the boundaries of accuracy. For instance, how prone is their finding that sectoral toroidal modes dominate to errors accrued during the inversion? We ask and answer that question here; this work also lays the foundation for future analyses of toroidal and eventually, poloidal flows in the Sun.

There are numerous inversion techniques with which to interpret seismic observations \citep[for a comparison between various inversion techniques, see][]{CD1990}, e.g., Regularized Least Squares \citep[RLS;][]{Jensen2001,Dombroski2013,MandH2018}, Multi-channel Deconvolution  \citep[MCD;][]{Jensen1998,Jensen2003,Zhao2003} and Subtractive Optimally Localized Averages \citep[SOLA:][]{PandT1992,Svanda2011,Jackiewicz2012}. In this paper, we focus on SOLA, applied to simulated mode-coupling measurements. SOLA algorithm entails finding a way to sum up the observables in such a manner as to allow an estimate of average value of convective flow at the desired depth (for more details, see section~\ref{inversions})

\subsection{Forward Model}\label{forwardmodel}
As convection is the main focus of this paper, we begin by using the Chandrashekhar-Kendall decomposition, or the Poloidal-Toroidal decomposition, that separates a vector field - here ${\bf u}_{o}({\bf r},\sigma)$, written as ${\bf u}_{o}^{\sigma}({\bf r})$ for compactness, where ${\bf u_0}$ is the vector convective flow, $r$ is the radial co-ordinate, $\sigma$ is the temporal frequency - into poloidal and toroidal components: 
\begin{equation}\label{flow}
\begin{aligned}
    {\bf u}_{o}^{\sigma}({\bf r}) = \sum_{s,t}\:u_{st}^{\sigma}(r)Y_{s}^{t}\hat{\bf{r}} \: +\; v_{st}^{\sigma}(r){\bf\nabla_{h}}Y_{s}^{t}\;\\ -\; i w_{st}^{\sigma}(r)\hat{\bf{r}}\times{\bf\nabla_h}Y_{s}^{t}.
\end{aligned}
\end{equation}
The subscripts $s$ and $t$ denote the angular degree and azimuthal order of the spherical-harmonic coefficients of the perturbation. The terms $u_{st}$, $v_{st}$ comprise poloidal flow and $w_{st}$ is the toroidal flow. The toroidal vector field, ${\bf T} = -i w_{st}^{\sigma}(r)\hat {\bf r}\times{\bf\nabla_h}Y_{s}^{t}$, is, by construction, mass conserving and only has lateral components of flow, i.e., $\hat {\bf r}\cdot {\bf T} =0$. The forward problem relates to connecting the flow to the observable which, in mode coupling, is the cross-correlation between line-of-sight oscillation wavefields at different spatio-temporal frequencies, i.e.  $\phi_{\ell' m'}^{\omega+\sigma}\phi_{\ell m}^{\omega*}$. The indices $\ell$, $\ell'$ and $m$, $m'$ are the angular degrees and the azimuthal wavenumbers of the wavefield respectively, $\sigma$ is the timescale associated with the perturbation, and $m'-m = t$ is the azimuthal wavenumber associated with the non-axisymmetric perturbation \citep{MandH2020}.  To simulate a frame that is co-rotating with the Sun, we replace $\sigma$ with $\sigma + t\Omega$ everywhere in the forward model, where $\Omega=453$nHz is the rotation rate at the equator \citep[e.g.][]{Wood2016}.

A crucial task in the forward model is to define a quantity called the flow sensitivity kernel that linearly connects changes in the observable $\phi_{\ell' m'}^{\omega+\sigma + t\Omega}\phi_{\ell m}^{\omega*}$ to changes in flow - the sensitivity kernel depends on oscillation mode indices ($n,\ell,m$) and ($n',\ell',m'$) with $n$ and $n'$ denoting the radial order, spherical harmonic wavenumbers ($s,t$) of the flow, and the radial dimension $r$. Throughout this paper, we consider only $n'=n$, $\ell'=\ell$ , also known as self-coupling; self-coupled modes are sensitive only to odd $s$ toroidal flow. We also set $m'=m+t$. We use an approximate form of the flow sensitivity kernel for the toroidal flow $w_{st}^{\sigma}(r)$ \citep[see][]{Vor2011, Wood2014} that is defined in Appendix \ref{appendixA}. Since it is more convenient to work with a condensed form the observable of $\phi_{\ell' m'}^{\omega+\sigma + t\Omega}\phi_{\ell m}^{\omega*}$, one that depends on the same variables as does the flow field $w_{st}^{\sigma}(r)$, namely $s,t$ and $\sigma$, we introduce (using the derivation in Appendix \ref{appendixA}) the $b$-coefficients ($b_{st}^{\sigma}(n, \ell)$) to establish the linear relation through sensitivity kernel as 
 \begin{equation}\label{bnoleak}
    b_{st}^{\sigma}(n, \ell) = f_{0,s}\int_{\odot}\; dr\; w_{st}^{\sigma}(r)\;\kappa_{n \ell}(r).
\end{equation}
where $f_{0,s}$ is a term that is defined to be non-zero only for odd $s$ and $\kappa_{n\ell}(r)$ is the term composed of radial and horizontal eigenfunctions for the ($n,\ell$) mode. The above equation also makes it clear that time-variation in the flow is captured in the measurement. With that in mind, we proceed with highlighting the forward model derived in \citet{Han2018}. To keep the notations of observations and model distinct, we denote the latter coefficients by $B$, which takes into account the effect of spatial leakage (Equation~(\ref{Bst})).

Because we do not observe the full Sun, we are unable to perfectly decompose and isolate the solar oscillation wavefield into its spherical-harmonic components. This results in a blurring of component peaks, i.e. power leaking from one spherical harmonic channel to its neighbours. \citet{SandB1994} modeled this effect of spatial leakage, quantified by the matrix $L_{\ell m}^{\ell' m'}$ in which a given element indicates the degree of leakage between two modes ($\ell,m$) and ($\ell',m'$). Using $L_{\ell m}^{\ell' m'}$, we incorporate spatial leakage into $B_{st}^{\sigma}(n,\ell)$, the $B$-coefficients obtained from mode-coupling model, as (contrast with Equation~(\ref{bnoleak}))
\begin{equation}\label{Bst}
    B_{st}^{\sigma}(n,\ell) = \sum_{s't'}\int_{\odot} dr\: w_{s't'}^{\sigma}(r)\Theta_{st}^{s't'}(r;n,\ell,\sigma),
\end{equation}
where the kernel $\Theta_{st}^{s't'}$ that encompasses oscillation mode spatial leakage linearly relates the $B$-coefficients to the flow $w_{st}^{\sigma}(r)$. The summation over $s'$ and $t'$ is indicative of mode leakage effect being translated into leakage in the flow field. (see Appendix \ref{appendixB} for more details).

\section{Outline of the work}\label{outline of the work}
Our goal is to determine the validity of SOLA as an inversion technique for normal-mode coupling applied to image convection. We first construct the $B_{st}^{\sigma}(n,\ell)$ using synthetics of toroidal flows by applying the forward model described by Equation~(\ref{Bst}). We then use the $B_{st}^{\sigma}(n,\ell)$ as the observables that need to be inverted to recover the average value of flow $w_{st}^{\sigma}(r_o)$ at different depths $r_o$ using SOLA.
Since we need a metric to assess the performance of the synthetic test, we compare the velocities  of the input and the recovered flow and if we obtain a good match between the two, we understand that as being a step towards validating SOLA for this problem.

Spatial leakage is an important issue to consider. Since the measurements ($B_{st}^{\sigma}$, that contains the information about correlation between different modes) is linearly related to the underlying perturbation ($w_{st}^{\sigma}(r)$), leakage in oscillation modes (from ($\ell, m$) to ($\ell', m'$)) implies that $B_{st}^{\sigma}$ contains signal from neighbouring channels ($s',t'$) in addition to its own ($s,t$) power. Therefore inversions that are carried out without taking this effect into account might lead to inaccurate (depending on the amount of leakage) inferences of the flow velocities. Hence we address this layer of complexity by penalizing the influx of power into the desired channel ($s,t$) from the neighbouring channels and see if we are able to isolate the desired ($s,t$) and the power it contains. We choose self-coupling, i.e., $\ell'=\ell$, because the amount of leakage between different oscillation modes is limited to the extent that we are able to model it in a relatively straightforward manner during inversion. Also, as discussed in section~(\ref{sec:intro}), validating inversions for toroidal flow as we do in this paper, albeit as a consequence of self-coupled modes, strengthens the foundation of mode-coupling as a method to study Rossby modes and differential rotation.

\subsection{Description of the synthetic test}
The angular degrees of the oscillation modes used in this work range from $70\leq \ell\leq 150$. The parameters required to calculate the frequency of a given mode $\omega_{n \ell m}$, the mode amplitude, captured by $N_{\ell}$, and the leakage matrix, $L_{\ell m}^{\ell' m'}$, are obtained from the Stanford data repository,
http://jsoc.stanford.edu/. 

\subsection{Synthetic flow details}\label{synthetic flow details}
We chiefly use as input two different simulations of convective flow used in \citet{HanSciA2020}, \citep[for more details, see][]{Hotta2016} and two other simulations as an independent means of validating our technique. The three variables common to all the four simulations are the angular degree $s$, which takes the range $1\leq s\leq 49$ (odd values only), azimuthal wavenumber $t$ spanning $-s\leq t\leq s$, with $t=0$ absent and temporal frequency $\sigma$ that spans the range $0.03\leq \sigma\leq 1.44$ (in $\mu$Hz), $45$ values in all. The four simulations are
\begin{enumerate}
    \item A 3D non-rotating convection simulation \citep{Hotta2016} in a spherical shell where the radial grid spans the range $0.71\leq r\leq 0.989$ (see Figures~\ref{fig_sim1} for results).
    \item A 3D convection calculation \citep{Hotta2016} with solar-like differential rotation in a spherical shell where the radial grid spans the range  $0.71\leq r\leq 0.959$ (see Figure~\ref{fig_sim2,3,4}, panels (a) and (b) for results).
    \item A velocity profile given by $w_{st}(r,\sigma)= 10^3 q s r$, where $q$ takes integer values in the range $1\leq q\leq 45$ (each of the 45 integers corresponds to a value of $\sigma$) and where the radial grid $r$ spans the range $0.71\leq r\leq 0.989$. A profile of such a nature was chosen to check if the synthetic test gives satisfactory results for any kind of input (see Figure~\ref{fig_sim2,3,4}, panels (c) and (d) for results).
    \item A velocity profile given by $w_{st}(r,\sigma)= re^{\iota\theta}$, where the radial grid spans the range $0.71\leq r\leq 0.989$ and $\theta$ is a uniform random number between [$0,2\pi$]. Randomness is introduced into the simulation to test the robustness of the technique in a different manner from the above three simulations (see Figure~\ref{fig_sim2,3,4}, panels (e) and (f) for results).
    \end{enumerate}

\section{Inversions}\label{inversions}
We consider the following three cases of inversions performed using SOLA namely 
\begin{enumerate}
    \item No leakage in the observables,
    \item Leakage in the observables without leakage penalty in the inversions,
    \item Leakage in the observables with leakage penalty in the inversions.
\end{enumerate}

\subsection{No leakage in the observables}\label{no leak}
We compute $b_{st}^{\sigma}(n,\ell)$ using Equation~(\ref{bnoleak}) and perform inversions. In SOLA, we try to combine our synthetic observables $b_{st}^{\sigma}(n,\ell)$ by finding a set of coefficients $c_{n\ell}$ such that the weighted sum of $b_{st}^{\sigma}(n,\ell)$ over ($n,\ell$) will give us an average value of flow $w_{st}^{\sigma}(r_0)$ around a desired depth $r_0$. That is, the value of flow at $r_0$ recovered from the inversion is
\begin{equation}\label{sumbst}
    w_{st}^{\sigma}(r_o) =  \sum_{n\ell}c_{n\ell}(r_o)b_{st}^{\sigma}(n,\ell).
\end{equation}
This is translated into an optimization problem by demanding an `averaging kernel', which is a weighted sum of the kernels $K_{n\ell}$ over ($n,\ell$), to resemble a well-localized function around $r_0$.
The averaging kernel, $\mathcal{K}(r,r_o)$, is defined as 
\begin{equation}\label{eqavgknoleak}
    \mathcal{K}(r,r_o) = \sum_{n\ell}c_{n\ell}(r_o)K_{n\ell}(r). 
\end{equation}
We desire the localization to be as sharp as allowable (as close to $\delta(r-r_o)$ as possible) by a sum of finite number of kernels \citep[see][for why sharp localization is at odds with noise minimization]{PandT1994}. For this purpose, the `target' kernel, $\mathcal{T}(r,r_{o})$, which is a template that we want our averaging kernel to match, is chosen as a Gaussian centered at $r_0$
\begin{equation}
    \mathcal{T}(r,r_o) = \frac{1}{\sqrt{2\pi}\Delta}exp\Big[-\frac{(r-r_0)^2}{2\Delta^2}\Big],
\end{equation}
whose width $\Delta$ can be chosen to be arbitrarily small in the absence of noise in the measurements \citep[see for e.g.,][section 2.1]{CD1990}.
Hence the optimization problem is posed in the matrix form \citep{PandT1994} as 
\begin{equation}
    A\:\{c\} = v,    
\end{equation}
where $c$ is the column vector of unknown, real $c_{n\ell}$. We solve this linear-algebra problem using Singular Value Decomposition,  with singular values $\epsilon$ cut-off at $\epsilon / \epsilon_{max} > 10^{-6}$. 
The matrix element $A_{n\ell,n'\ell'}$ is given by 
\begin{equation}
    A_{n\ell,n'\ell'} = \int_{\odot}\;dr\;K_{n\ell}(r)K_{n'\ell'}(r),
\end{equation}
and 
\begin{equation}
    v = \int_{\odot}\;dr  \:K_{n\ell}(r)\mathcal{T}(r,r_o).
\end{equation}
The true value of flow can then be obtained at $r_0$ by
\begin{equation}\label{trueflow}
    w_{st}^{\sigma}(r_o) = \int_{\odot}\;dr\;\mathcal{T}(r,r_o)  w_{st}^{\sigma}(r). 
\end{equation} 
The integral over radius with the target function is in keeping with the spirit of estimating an average value of the flow around $r_o$ similar to Equation~(\ref{sumbst}) where we combine our observables $b_{st}^{\sigma}$ by taking their weighted sum with $c_{n\ell}$ in order to obtain an average (the $b_{st}^{\sigma}$ themselves are in a sense an averaging quantity since it is an integral over the radius; see Equations~(\ref{bnoleak}) and~(\ref{Bst})).

\subsection{Leakage in the observables and no penalty}\label{leakage no penalty}
 The matrix $A$ and the R.H.S. $v$ of the matrix problem are rewritten using the full kernel from Equation~(\ref{fullkernel}) as
\begin{equation}
 A_{n\ell,n'\ell'} = \int_{\odot} \:dr \:\Theta_{st}^{st}(r;n,\ell,\sigma)\:\Theta_{st}^{st}(r;n',\ell',\sigma).
\end{equation} 
and 
\begin{equation}
    v = \int_{\odot} \:dr \:\Theta_{st}^{st}(r;n,\ell,\sigma)\:\mathcal{T}(r,r_0).
\end{equation}
We redefine the the averaging kernel as 
\begin{equation}\label{avgkleak}
    T_{st}^{s't'}(r,r_o) = \sum_{n\ell}c_{n\ell}(r_o)\Theta_{st}^{s't'}(r;n,\ell,\sigma),
\end{equation} 
and use $B_{st}^{\sigma}(n,\ell)$ instead of $b_{st}^{\sigma}(n,\ell)$ wherever applicable; for e.g., Equation~(\ref{sumbst}) is rewritten after including leakage as

\begin{equation}\label{sumBst}
    w_{st}^{\sigma}(r_o) \approx  \sum_{n\ell}c_{n\ell}(r_o)B_{st}^{\sigma}(n,\ell).
\end{equation}
The '$\approx$' symbol implies that since the $B_{st}^{\sigma}(n,\ell)$ contains power leaked from neighbouring channels $s'$ and $t'$ as seen from Equation~(\ref{Bst}), the $w_{st}^{\sigma}(r_o)$ can only be an approximate quantity of the average value of flow at $r_o$. Also, ideally, since we compute $\Theta_{st}^{st}(r;n,\ell,\sigma)$ at all the values $\sigma$ highlighted in section~(\ref{synthetic flow details}), it would follow that the inversions also be carried out at all the values of $\sigma$. We instead consider the kernels $\Theta_{st}^{st}(r;n,\ell,\sigma)$ at an average value of $\sigma=1\mu$Hz throughout the inversion procedure and in Figure~\ref{fig_invcomp}, we show a difference in the results, albeit minor, when using only one value of $\sigma$ versus using all the values of $\sigma$.

\subsection{Penalizing the leaked power}\label{penalizing the leaked power}
As described in section \ref{outline of the work}, we attempt to mitigate mode leakage, i.e., we try to diminish the contributions from ($s',t'$) to ($s,t$) by rewriting $A$ matrix as \citep[see][equation 35]{Han2018}
\begin{equation}\label{penalized A matrix}
\begin{aligned}
A_{n\ell,n'\ell'} = &\ \int_{\odot} \:dr \: \Big\{ \Theta_{st}^{st}(r;n,\ell,\sigma)\:\Theta_{st}^{st}(r;n',\ell',\sigma)\: +\\ &\  \lambda\sum_{s',t'}\:\Theta_{st}^{s't'}(r;n,\ell,\sigma)\:\Theta_{st}^{s't'}(r;n',\ell',\sigma)\Big\},
\end{aligned}
\end{equation}
so as to minimize the power leak from the neighbouring ($s,t$), suggested by the second term on the R.H.S..
The choice of $\lambda$ is obtained through trial and error and the rest of the inversion procedure follows the same as before.

Although \citet{PandT1992} point out that OLA - Optimally Localized Averages - is more popular than Regularized Least-Squares (RLS; refer Appendix~\ref{appendixC} for details) since the former produces more highly localized averaging kernels and are hence easier to interpret, they still consider SOLA to be superior. This is due to two reasons - one, we are able to curate the target form of the averaging kernel suitable to the presence / absence of noise, and two, the amount of computation (number of matrix inversions) is reduced by a factor equal to the number of radii at which inversions are performed to obtain an estimate of the flow.

\subsection{Defining metric for comparison}\label{defining metric for comparison}
Velocities obtained from the flow power-spectrum are a useful means of comparing models of turbulence \citep[see][]{YandM}. Considering that it is cumbersome to make comparisons for all the $s$, $t$, and $\sigma$ at various depths, we instead compute three different spectral averages to compare between true and recovered flow. To be consistent with H20, we use the definition of power-spectrum $P_{st}^{\sigma}(r) = s(s+1)|w_{st}^{\sigma}(r)|^2$ (where the derivation for the factor $s(s+1)$ can be found in the supplementary materials of H20).
\begin{equation}\label{eqPs}
    P(s,r_0) = \sum\limits_{t,\sigma}P_{st}^{\sigma}(r_o) = \sum\limits_{t,\sigma}s(s+1)|w_{st}^{\sigma}(r_o)|^2
\end{equation}
\begin{equation}\label{eqPsig}
    P^{\sigma}(r_0) = \sum\limits_{s,t}P_{st}^{\sigma}(r_o) = \sum\limits_{s,t}s(s+1)|w_{st}^{\sigma}(r_o)|^2
\end{equation}
\begin{equation}\label{eqPdiff}
    P(s-|t|,r_0) = \sum\limits_{\sigma}P_{st}^{\sigma}(r_o) = \sum\limits_{\sigma}s(s+1)|w_{st}^{\sigma}(r_o)|^2
\end{equation}

$P(s)$ is a useful quantity for highlighting convective velocity and length scales and to understand if the simulations of convection used in the synthetic test show the same trend in power variation with $s$ as observations - for instance, simulations show velocity decreasing with increasing $s$, (see Figure~\ref{fig_sim1}), whereas observations show increasing power with increasing $s$  \citep[see][for more details]{Han2016,HanSciA2020}.

$P(\sigma)$ characterizes the variation of power with temporal frequency channels. Highlighted peaks in frequency are indicative of special temporal scales of toroidal flow evolution.

$P(s-|t|)$ sheds light on the shape of convection by characterizing the distribution of power in sectoral, tesseral and zonal modes.

The above definitions of the velocities ride on the simplistic assumption $P_{st}^{\sigma}(r_0) = s(s+1)|w_{st}^{\sigma}(r_o)|^2$. Using Equation~(\ref{sumBst}) for our simple definition of $P_{st}^{\sigma}(r_0)$, we have 
\begin{align}\label{eq1}
\begin{split}
P_{st}^{\sigma}(r_0)  = \sum_{n\ell}\sum_{n'\ell'}\;c_{n\ell}(r_o)c_{n'\ell'}(r_o) B_{st}^{\sigma}(n,\ell)B_{st}^{\sigma*}(n',\ell') = \sum_{s',t',s{''},t{''},n,\ell,n',\ell'} f^\sigma(...)\,w^\sigma_{s't'}\,w^{\sigma*}_{s{''}t{''}},
\end{split}
\end{align}
where we have substituted the expression connecting the $B$ coefficients and flows in Equation~(\ref{Bst}) ; the term $f^\sigma(...)$ is a product of the kernels evaluated at the two sets of radial orders, mode orders and spherical harmonic wavenumbers of the perturbation. Relevant to the discussion at hand are cross products of flow terms $w_{st}^{\sigma}$. This implies that the power content in a given $s,t$ channel is not fully isolated and contains signals from neighbouring $s',t'$ channels. As different flow models $w_{st}^{\sigma}$ have differing covariance matrices, i.e. ranging from fully correlated to entirely uncorrelated, the inferred velocities from the inversions that do not model the covariance may pull up short of expectations in terms of accuracy, as shown in Figure~\ref{fig_sim2,3,4}, panels (c) through (f).

To reiterate, the steps for the synthetic test are as follows.
\begin{enumerate}
    \item We use the forward model (section~(\ref{forwardmodel})) to compute the $B$-coefficients using the flow models described in section~(\ref{synthetic flow details}), without leakage (Equation~(\ref{bnoleak})) and with leakage (Equation~(\ref{Bst})).
    \item We treat these synthetically generated $B$-coefficients as `observations' and solve the inverse problem using SOLA to obtain the  an estimate of flow at the depth $r_o$. We also demonstrate the improvement in results by penalizing the contribution of power from the neighbouring modes to the desired mode as described in section~(\ref{penalizing the leaked power}).
    \item The power-spectrum $P_{st}^{\sigma}(r) = s(s+1)|w_{st}^{\sigma}(r)|^2$ is calculated and averaged over combinations of variables $s,t,\sigma$ (described in equations~(\ref{eqPs}) through~(\ref{eqPdiff})) to compress information. To keep the results consistent with H20, we plot comparisons of the averages of the velocities, $\sqrt{P_{st}^{\sigma}(r)}=\sqrt{s(s+1)|w_{st}^{\sigma}(r)|^2}$.
\end{enumerate}

\section{Results and discussion}\label{results}
\begin{figure}
\gridline{\fig{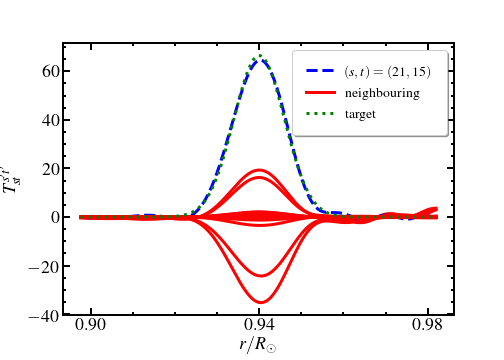}{0.48\textwidth}{(a)}
          \fig{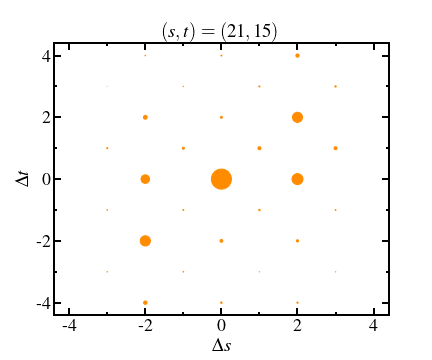}{0.41\textwidth}{(b)}
          }
\gridline{\fig{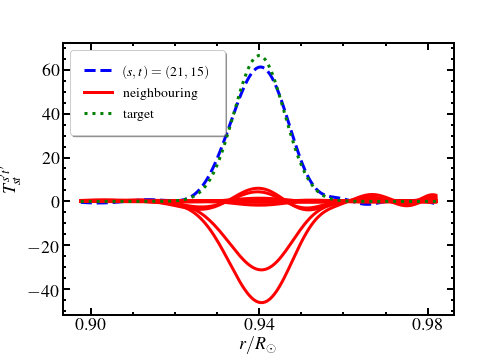}{0.48\textwidth}{(c)}
          \fig{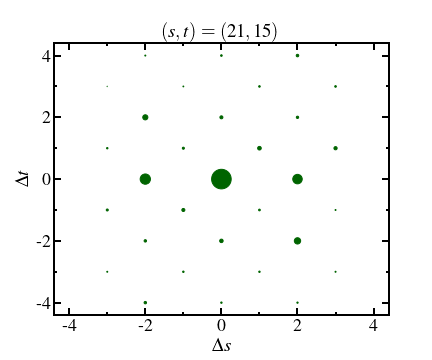}{0.41\textwidth}{(d)}
          }          
\caption{(a): Target Gaussian, averaging kernel of the desired wavenumber $(s,t)=(21,15)$, centered at $r=\;0.94R_{\odot}$, and averaging kernels of the neighbouring $(s',t')$, with no leakage penalties. (b): Size of the dots indicate the amount of leakage i.e., the amount of power present in $(s,t)=(21,15)$ and each of the neighbouring $(s',t')$. (c) and (d): Same as (a) and (b), but with leakage penalties. Power from some of the neighbouring $(s',t')$ has been suppressed after the inclusion of penalty.
\label{fig_avgk2115}}
\end{figure}

\begin{figure}
\gridline{\fig{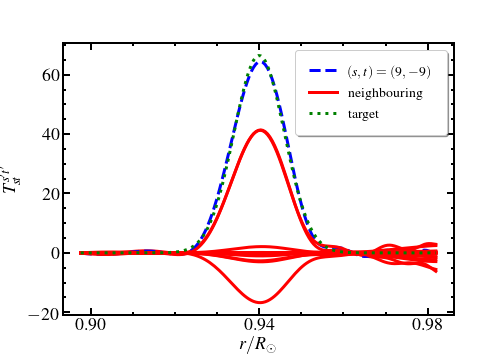}{0.48\textwidth}{(a)}
          \fig{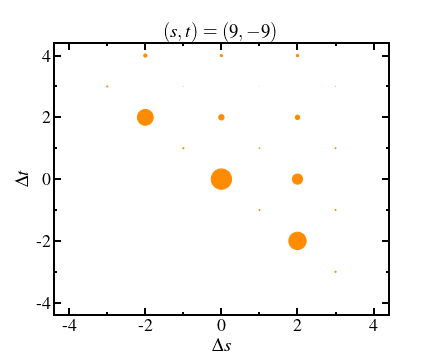}{0.41\textwidth}{(b)}
          }
\gridline{\fig{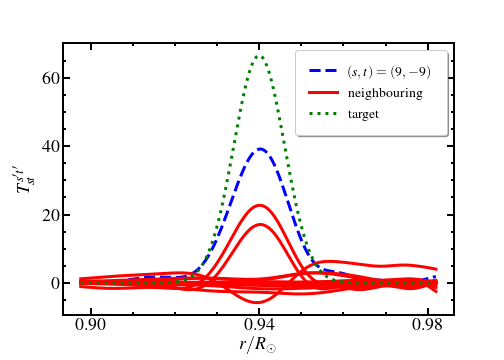}{0.48\textwidth}{(c)}
          \fig{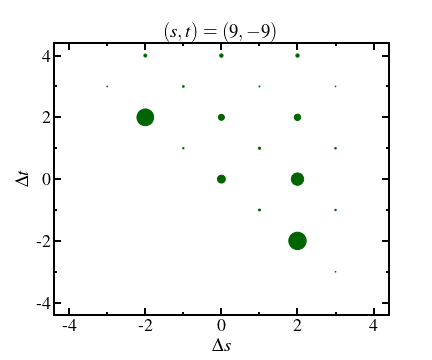}{0.41\textwidth}{(d)}
          }          
\caption{(a): Target Gaussian, averaging kernel of the desired wavenumber $(s,t)=(9,9)$, centered at $r=\;0.94R_{\odot}$, and averaging kernels of the neighbouring $(s',t')$, with no leakage penalties. (b): Size of the dots indicate the amount of leakage i.e., the amount of power present in $(s,t)=(9,9)$ and each of the neighbouring $(s',t')$. (c) and (d): Same as (a) and (b), but with leakage penalties. The penalty did not work as expected, suppressing the power in the desired ($s,t$) and raising it in the neighbouring ($s',t'$).
\label{fig_avgk99}}
\end{figure}

We show comparisons for various averages of the velocities defined in section~(\ref{defining metric for comparison}) for all the simulations. There are four curves shown in each figure - the black curve is the true power-spectrum or the input given by Equation~(\ref{trueflow}), the other three curves given by Equation~(\ref{sumbst}), Equation~(\ref{sumBst}) and Equation~(\ref{sumBst}) with $c_{n\ell}$ modified to include penalty. We find that, in all these figures, the inversions for the case mentioned in section~(\ref{no leak}), wherein there is no leakage present in the observables (Equation~(\ref{bnoleak})), is a perfect match with the input. We also show figures for the averaging kernels, without and with penalty for different ($s,t$) in Figures~\ref{fig_avgk2115} and~\ref{fig_avgk99}.

\begin{figure}
\gridline{\fig{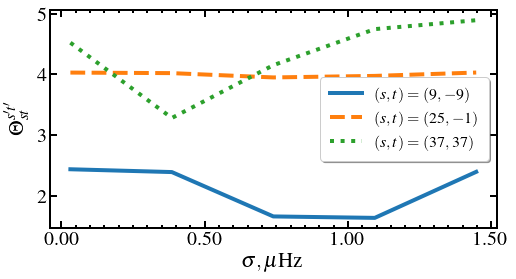}{0.48\textwidth}{(a)}
          \fig{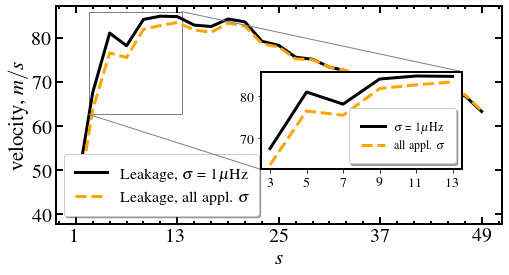}{0.48\textwidth}{(b)}
          }
\caption{(a): The kernel coefficient $\Theta_{st}^{s't'}/\kappa_{n\ell}(r)$ from Equation~(\ref{fullkernel}) for $(n,\ell)=(2,70)$, where the division by $\kappa_{n\ell}(r)$ removes the radial dependence. Although the kernel coefficient changes with frequency $\sigma$ and across different ($s,t$), the percentage variation with respect to the average value at $\sigma=1\mu$Hz is small enough to permit us to assume that it is invariant with $\sigma$. Invoking this assumption allows for performing inversions at comparable accuracy with much lower computational cost. (b): Using the simulation (1), we compare velocity $\sqrt{s(s+1)\sum\limits_{t,\sigma} |w_{st}^{\sigma}(r_0)|^2}$ described in equation~(\ref{eqPs}) when using $\sigma=1\mu$Hz as compared to all the values of $\sigma$ highlighted in section~(\ref{synthetic flow details}).
    \label{fig_invcomp}}
\end{figure}

\begin{figure}
\gridline{\fig{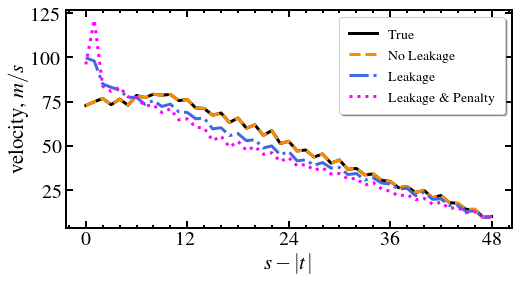}{0.48\textwidth}{(a), $r=\;0.98R_{\odot}$}
          \fig{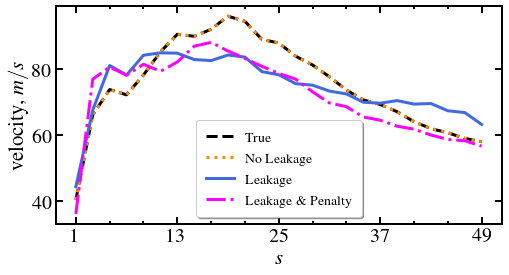}{0.48\textwidth}{(b), $r=\;0.98R_{\odot}$}
          }
\gridline{\fig{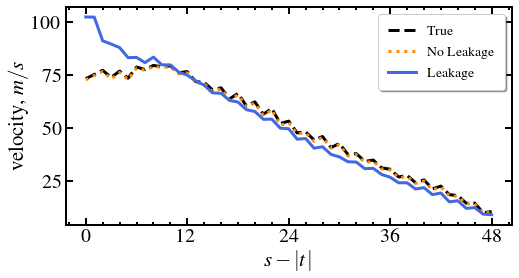}{0.48\textwidth}{(c), $r=\;0.98R_{\odot}$}
          \fig{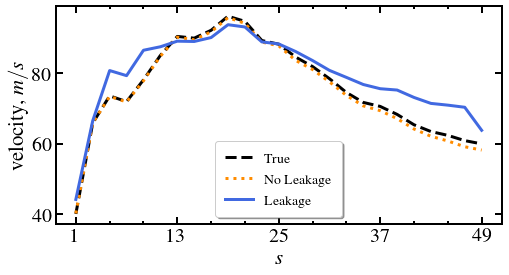}{0.48\textwidth}{(d), $r=\;0.98R_{\odot}$}
          }          
\caption{\textit{Results for simulation 1, section~(\ref{synthetic flow details}}).  (a): Velocity computed using equation~(\ref{eqPdiff}). Note that velocity declines as $s-|t|$ increases, implying more power in sectoral modes than in tesseral and zonal modes. (b): Velocity computed using equation~(\ref{eqPs}). Velocity peaks around $s\approx19$ but declines thereafter. This is contrary to the trend seen in observations (e.g. H20). Panels (c) and (d): same computations as in (a) and (b), using RLS as described in Appendix~\ref{appendixC}. RLS overestimates velocities at lower $s-|t|$ as compared to SOLA, but provides a better match than SOLA for velocity variation with $s$.}
\label{fig_sim1}
\end{figure}

Figure~\ref{fig_invcomp} demonstrates that significant computational savings are obtained by considering the kernels $\Theta_{st}^{s't'}(r;n,\ell,\sigma)$ at one average value, i.e., a single frequency ($\sigma = 1\mu$Hz in this work) during the inversion. Figure~\ref{fig_sim1}, panels (a) and (c) (for simulation (1), section~(\ref{synthetic flow details})) shows that inversions using SOLA  perform better than RLS in understanding the geometry of the convective features (equation~(\ref{eqPdiff})). While leakage implies that power is redistributed across modes, careful optimization in the inversion can reliably mitigate this, thereby reducing the errors in inferring power. Figure~\ref{fig_sim2,3,4}, panel(a) (results for simulation (2), section~(\ref{synthetic flow details})) paints almost the same picture as Figure~\ref{fig_sim1} in that the inversions work well in reproducing the output velocity. However, in panel (b), the penalty makes virtually no difference. Figure~\ref{fig_sim2,3,4}, panels (c) through (f) show a stark difference in the quality of results that is produced when flow $w_{st}^{\sigma}(r)$ is chosen to be a simple function of its dependent variables (simulations (3) and (4), section~(\ref{synthetic flow details})). We reiterate that different flow models have different covariance matrices, ranging from fully correlated to fully random, comprising a number of unknowns that need to be taken into account in order to obtain accurate inversions. Nonetheless, the velocity dependence on angular wavenumber is reproduced even if there is a poor match in amplitude. It is also useful to determine how to mitigate leakage-related errors uniformly while keeping in mind that we can only curate specific parameters (e.g., $\lambda$ in Equation~(\ref{penalized A matrix})) involved in the inversion algorithm.

Figure~\ref{fig_avgk2115} shows the averaging kernels, $T_{st}^{s't'}(r)$ defined in Equation~(\ref{avgkleak}), for $(s,t)=(21,15)$ without (panel(a) and (b)) and with penalty (panel (c) and (d)), the latter showing the improvement in the inference of the desired ($s,t$) by diminishing the contribution from the neighbouring ($s',t'$). Figure~\ref{fig_avgk99} makes it clear as to why errors in the inversions may be attributed to the imperfect penalization for sectoral modes (see Figure~\ref{fig_sim2,3,4}, panel (c) and (e) - a substantial amount of power is concentrated in the sectoral modes).

\begin{figure}
\gridline{\fig{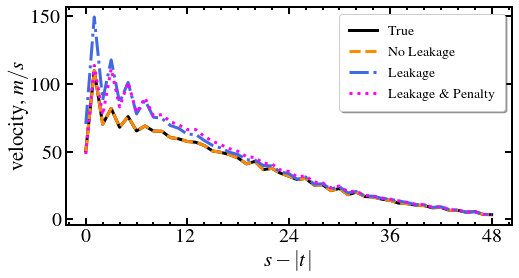}{0.48\textwidth}{(a), $r=\;0.90R_{\odot}$}
          \fig{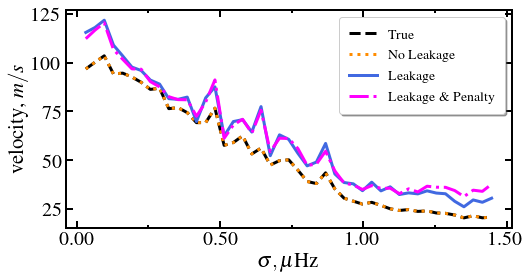}{0.48\textwidth}{(b), $r=\;0.94R_{\odot}$}
          }
\gridline{\fig{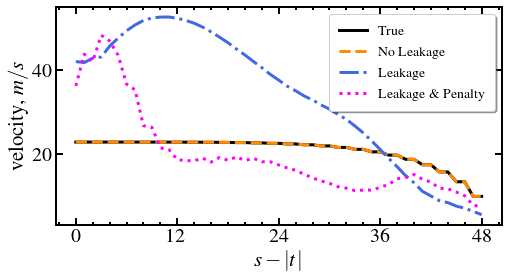}{0.48\textwidth}{(c), $r=\;0.94R_{\odot}$}
          \fig{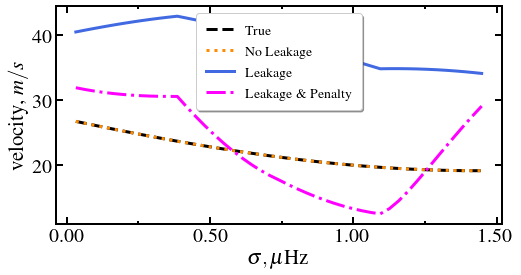}{0.48\textwidth}{(d), $r=\;0.94R_{\odot}$}
          }  
\gridline{\fig{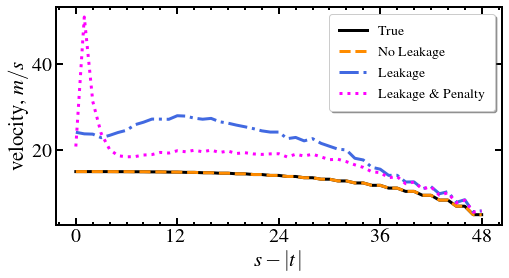}{0.48\textwidth}{(e), $r=\;0.98R_{\odot}$}
          \fig{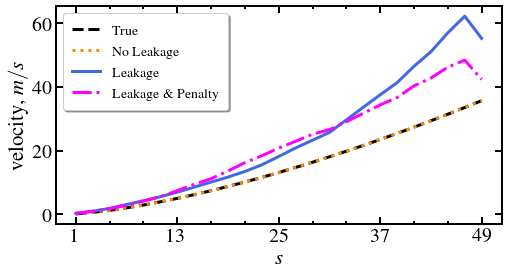}{0.48\textwidth}{(f), $r=\;0.98R_{\odot}$}
          }  
\caption{\textit{Results for simulation 2 in panels (a) and (b), simulation 3 in panels (c) and (d), and simulation 4 in panels (e) and (f), section~(\ref{synthetic flow details})}. Panel (a): Sectoral modes are more dominant than tesseral and zonal modes. Panel (b): velocity computed using equation~(\ref{eqPsig}). The peaks at certain frequencies denote the overall evolution period of the toroidal flow $w_{st}^{\sigma}(r)$.
(c): Leakage in the observables and the inversion causes the algorithm to overestimate the velocity. Penalizing the contribution from the neighboring modes can mitigate leakage but the results are seen to be unreliable. This is because the covariance model for $w^\sigma_{st}$ is not accurately taken into account.} (d): Velocity is systematically over-estimated at all values of $\sigma$. Although penalty mitigates leakage, imperfect optimization leads to a poor match between the recovered and true velocity.
(e): Velocity is systematically over-estimated at all values of $s-|t|$. Penalty makes the results worse for sectoral modes ($s-|t|\sim0$) as is evident from Figure~\ref{fig_avgk99}. (f): Velocity is systematically over-estimated at all values of $s$ despite penalty.
\label{fig_sim2,3,4}
\end{figure}

\section{Conclusion}
In this work, we made use of four simulations of toroidal convective flow, treated them as input to the mode-coupling model, and constructed realistic observables. We then used the inversion algorithm SOLA to invert the observables and to obtain an average value of the flow at different depths. Using another inversion algorithm - RLS - we find that results are comparable with SOLA, but we prefer SOLA since the concept of averaging kernel is well defined (equations~(\ref{eqavgknoleak}) and~(\ref{avgkleak}) and Figures~\ref{fig_avgk2115} and~\ref{fig_avgk99}). We also tried to address the undesired effect of spatial leakage (an effect which is rather pronounced in Figure~\ref{fig_sim2,3,4}, bottom four panels) by leveraging a penalty term in the optimization problem in the inversion. We were able to successfully recover the input for simulations (1) and (2) but the results were unsatisfactory for simulations (3) and (4). This leads us to believe that the particular model of the flow chosen as input dictates how well the synthetic test can perform since to fully understand how the flow behaves across all its dependent variables $s$, $t$, $\sigma$ and $r$, we need to be able to write down a covariance matrix $w_{st}^{\sigma}(r)w_{s't'}^{\sigma'*}(r')$ of the convective flow itself. A synthetic test that discounts this effort exposes itself to limitations in being able to accurately recover the input. But we stress that it is critical to test inversions using synthetics in order to improve faith in observational inferences. This paper lays the groundwork for normal-mode coupling as a successful measurement tool to investigate the dynamics of the solar interior as we demonstrate that the synthetic measurements constructed using mode-coupling theory may be inverted to recover the perturbation using SOLA. 
Inversions of global-mode time-series of Michelson Doppler Imager (MDI) and Global Oscillations Network Group (GONG), utilizing much of the same legwork endured in this project, would be the logical next step in advancing mode-coupling as a useful method in being able to acquire a greater understanding of solar convection.

\appendix
\section{Relation between wavefield correlation and coupling coefficients}\label{appendixA}
The terms in the asymptotic form of the toroidal flow kernel 
\begin{equation}\label{Knlappendix}
    K_{st}(n,\ell;r) \simeq f_{0,s}\gamma^{\ell s \ell}_{tm}\;\kappa_{n\ell}(r),
\end{equation}
are given by
\begin{equation}
    \kappa_{n,\ell}(r) = \frac{(-1)^{\ell}}{\sqrt{2\pi}}\ell^{\frac{3}{2}}[ U_{n\ell}^{2} + \ell(\ell+1)V_{n\ell}^{2}],    
\end{equation}
where $U_{n\ell}$ and $V_{n\ell}$ are  the radial and horizontal eigenfunctions for the ($n,\ell$) mode and
\begin{equation}
    f_{\ell' - \ell , s} = (-1)^{\frac{s+\ell'-\ell-1}{2}}\frac{(s+\ell'-\ell)!!(s+\ell-\ell')!!}{\sqrt{(s+\ell'-\ell)!(s+\ell-\ell')!}},
\end{equation}
and
\begin{equation}\label{wigner}
    \gamma^{\ell ' s \ell}_{tm} = (-1)^{m+t} \;\sqrt{2s+1}\;   \begin{pmatrix}
\ell'& s &\ell\\
-(m+t)& t& m
\end{pmatrix}.
\end{equation}
The RHS. of Equation~(\ref{wigner}) is a modified Wigner-$3j$ symbol. The conditions $|m+t|\leq\ell'$, $|m|\leq\ell$, $|t|\leq s$, $|\ell'-\ell|\leq s$, $|\ell'-s|\leq \ell$ and $|\ell-s|\leq \ell'$ have to be met for the R.H.S. to be non-zero.
These self-coupled modes ($\ell'=\ell$) are sensitive only to odd-degree toroidal flow \citep{LandR1992}, $w_{st}^{\sigma}(r)$. This allows us to connect wavefield correlation to the flow as
\begin{equation}
    \phi_{\ell m+t}^{\omega+\sigma+t\Omega}\phi_{\ell m}^{\omega*} = H_{\ell\ell mt}^{\sigma}(\omega)\sum_{s}\gamma_{tm}^{\ell s\ell}\int_{\odot}\;dr\;f_{0,s}w_{st}^{\sigma}(r)\kappa_{n\ell}(r).
\end{equation}
where $H_{\ell \ell m t}^{\sigma}(\omega)$ is defined in Equation~(\ref{Hfunc}).
As suggested by \citet{Wood2016}, we use as the observable a linear-least-square fits approximation, $b_{st}^{\sigma}(n \ell)$, to the raw wavefield correlation $\phi_{\ell m+t}^{\omega+\sigma+t\Omega}\phi_{\ell m}^{\omega*}$, given by the reciprocal relations
\begin{equation}\label{phiphi*b}
    \phi_{\ell m+t}^{\omega+\sigma+t\Omega}\phi_{\ell m}^{\omega*} = \sum_{s}\gamma_{tm}^{\ell s \ell}H_{\ell\ell m t}(\omega)^{\sigma}b_{st}^{\sigma}(n,\ell),
\end{equation}
and
\begin{equation}\label{bphiphi*}
    b_{st}^{\sigma}(n, \ell) = \frac{\sum\limits_{m, \omega}\gamma^{\ell s\ell}_{tm}H_{\ell \ell m t}^{\sigma*}(\omega)\phi_{\ell m+t}^{\omega+\sigma + t\Omega}\phi_{\ell m}^{\omega*}}{\sum\limits_{m, \omega}|H_{\ell \ell m t}^{\sigma}(\omega)\gamma^{\ell s\ell}_{tm}|^2}.
\end{equation}
The $b_{st}^{\sigma}(n, \ell)$, known as $b$-coefficients, are a more amenable quantity than the $\phi_{\ell' m+t}^{\omega+\sigma+t\Omega}\phi_{\ell m}^{\omega*}$ since the former condenses all the $\omega$ and $m$ samples of the correlation.

\section{Full kernel}\label{appendixB}
The spectral profile of a mode \citep{And1990} is given by 
\begin{equation}
    R_{\ell m}^{\omega}=\frac{1}{(\omega_{n \ell m} - \iota \Gamma_{n\ell}/2)^2-\omega^2}.
\end{equation}
The addition of a small imaginary component $\Gamma_{n\ell}$ to the central frequency, $\omega_{n \ell m}$, captures the damping of the mode ($n,\ell$).
This helps in defining $H_{\ell\ell' m t}^{\sigma}$, a weighting function, (see Eq.~[\ref{phiphi*b}] and Eq.~[\ref{bphiphi*}]), that includes Lorentzians associated with wavenumbers ($\ell, m$) and ($\ell',m'$) along with their normalization constants $N_{\ell}$ and $N_{\ell'}$ \citep[see for e.g.,][Appendix C]{Han2018}, according to
\begin{equation}\label{Hfunc}
    H_{\ell \ell' m t}^{\sigma}(\omega) = -2\omega(N_{\ell'}\:R_{\ell m}^{\omega*}\:|R_{\ell' m'}^{\omega+\sigma + t\Omega}|^{2}\:+\:N_{\ell}\:R_{\ell' m'}^{\omega+\sigma + t\Omega}\:|R_{\ell m}^{\omega}|^{2}).
\end{equation}
This allows us to now write $B_{st}^{\sigma}(n,\ell)$, the $B$-coefficients obtained from mode-coupling model, as (contrast with Equation~(\ref{bnoleak}))
\begin{equation}\label{Bstappendix}
    B_{st}^{\sigma}(n,\ell) = \sum_{s't'}\int_{\odot} dr\: w_{s't'}^{\sigma}(r)\Theta_{st}^{s't'}(r;n,\ell,\sigma),
\end{equation}
The kernel $\Theta_{st}^{s't'}$ that incorporates spatial leakage is given by
\begin{equation}\label{fullkernel}
\begin{aligned}
    \Theta_{st}^{s't'}(r;n,\ell,\sigma) = &\ N_{\ell st}^{\sigma}\kappa_{n\ell}(r)\sum_{\ell', \ell", m, m', \omega}f_{\ell"-\ell',s'}\:L_{\ell m}^{\ell' m'}\times \\
   &\ L_{\ell m+t}^{\ell" m'+t'}\:\gamma_{tm}^{\ell s\ell}\:L_{\ell m}^{\ell m}\: L_{\ell m+t}^{\ell m+t}\:H_{\ell \ell m t}^{\sigma + t\Omega*}(\omega)\;\times\\ &\gamma_{t'm'}^{\ell"s'\ell'}
   \;H_{\ell'\ell"m't'}^{\sigma + t\Omega}(\omega),
\end{aligned}
\end{equation}
with
\begin{equation}
    N_{\ell s t} = \frac{1}{\sum\limits_{m,\omega}|L_{\ell m}^{\ell m}\;L_{\ell m+t}^{\ell m+t}\;H_{\ell\ell m t}^{\sigma + t\Omega}(\omega)\;\gamma_{tm}^{\ell s\ell}|^2}.
\end{equation}

For the summation interval over $\omega$, we use one line-width from the resonance as 

$\omega\: \large\epsilon \: (\omega_{n\ell m} - \Gamma_{n\ell} ,\:\omega_{n \ell m} + \Gamma_{n\ell})$ or 
 $\omega\: \large\epsilon \: (\omega_{n\ell m+t} - \Gamma_{n\ell} + \sigma + t\Omega,\:\omega_{n \ell m+t} + \Gamma_{n\ell}+\sigma + t\Omega)$.

The leakage element decreases as $|\ell'-\ell|$, $|\ell''-\ell|$, $|s'-s|$ increase, $H_{\ell \ell m t}$, $H_{\ell' \ell'' m' t'}$ due to the twin effects of the Lorentzian profile rapidly decaying away from resonance and the leakage matrices diminishing in magnitude. Hence, we use the range $|\ell'-\ell|,|\ell''-\ell|\leq 2$, $|s'-s|\leq3$ and $|t'-t|\leq4$.

\section{RLS}\label{appendixC}
Using the same notation observed in \citet{MandH2020} section $3.2$, we decompose the flow $w_{st}^{\sigma}(r)$ in $B$-spline basis as
\begin{equation}\label{wstbspline}
    w_{st}^{\sigma}(r) = \sum_{k}\beta^{k\sigma}_{st}B_{k}(r),
\end{equation}
where $B_{k}$ is a $B$-spline basis function of order $3$. The relation between the total number of $B$-spline basis functions (i.e., the total number of $k$) and the number of knots is given by 
\begin{equation}
    \#k = \#knots - order -1
\end{equation}
Choosing $51$ knots, we get \#$k=47$. 
\subsection{No leakage in the observables}
We try to determine the coefficients $\beta^{k\sigma}_{st}$ by minimizing the misfit
\begin{equation}
    \chi = \sum_{n,\ell}\Big(b_{st}^{\sigma}(n,\ell)-f_{0,s}\int_{\odot} dr\: w_{st}^{\sigma}(r)\kappa_{n\ell}(r)\Big)^2 + \lambda\left(\frac{d^2 w_{st}^{\sigma}(r)}{dr^2}\right)^2,
\end{equation}
where the second term in the R.H.S. is a second derivative smoothing with $\lambda$ as the regularization parameter. The second term is optional and we do not make use of it in our inversions. The problem is turned into solving the system of equations
\begin{equation}\label{matrixeq1}
\sum_{k}f_{0,s}\int_{\odot} dr\:B_{k}(r)\;\kappa_{n\ell}(r)\;\beta^{k\sigma}_{st}=  b_{st}^{\sigma}(n,\ell).
\end{equation}
Setting 
\begin{equation}
    F_{st}^{k\sigma}(n,\ell) = f_{0,s}\int_{\odot} dr\:B_{k}(r)\;\kappa_{n\ell}(r),
\end{equation}
in Equation~(\ref{matrixeq1}), we have 
\begin{equation}\label{matrixeq1.1}
    \sum_{k}\:F_{st}^{k\sigma}(n,\ell)\;\beta^{k\sigma}_{st} = b_{st}^{\sigma}(n,\ell).
\end{equation}
Writing the matrix form of Equation~(\ref{matrixeq1.1}) as
\begin{equation}\label{Matrixform}
    F\:\{\beta\}=b_{st}^{\sigma}(n,\ell).
\end{equation}
Dropping the $s,t,\sigma$ for simplicity in notation, $F$ is given by (661 x 47)
\begin{equation}
\begin{pmatrix}
F^{k_{1}}(n_{1},\ell_{1}) & F^{k_{2}}(n_{1},\ell_{1}) & \cdots & F^{k_{47}}(n_{1},\ell_{1}) \\
F^{k_{1}}(n_{2},\ell_{2}) & F^{k_{2}}(n_{2},\ell_{2}) & \cdots & F^{k_{47}}(n_{2},\ell_{2}) \\
\vdots  & \vdots  & \ddots & \vdots  \\
F^{k_{1}}(n_{661},\ell_{661}) & F^{k_{2}}(n_{661},\ell_{661}) & \cdots & F^{k_{47}}(n_{661},\ell_{661}) 
\end{pmatrix},
\end{equation}
and $\beta$ is the unknown column vector to be determined, given by (47 x 1)
\begin{align}
        \begin{pmatrix}
           \beta^{k_{1}} \\
           \beta^{k_{2}} \\
           \vdots \\
           \beta^{k_{47}}
        \end{pmatrix}.
\end{align}
The R.H.S. is given by (661 x 1)
\begin{align}
    \begin{pmatrix}
           b_{st}^{\sigma}(n_1,\ell_1) \\
           b_{st}^{\sigma}(n_2,\ell_2) \\
           \vdots \\
           b_{st}^{\sigma}(n_{661},\ell_{661})
         \end{pmatrix}
\end{align}
\subsection{Leakage in the observables}
The entire procedure remains the same except we substitute 
\begin{equation}
    b_{st}^{\sigma}(n,\ell) \longrightarrow B_{st}^{\sigma}(n,\ell)
\end{equation}
and use only self-leakage, i.e., $(s',t')=(s,t)$
\begin{equation}
    f_{0,s}\kappa_{n\ell}(r) \longrightarrow \Theta_{st}^{st}(r;n,\ell,\sigma)
\end{equation}
with $\Theta_{st}^{st}(r;n,\ell,\sigma)$ evaluated at $\sigma=1\mu$Hz only, similar to SOLA, so as to reduce computation time.

\end{document}